\documentclass[superscriptaddress,showpacs,aps,pra,twocolumn,amsmath,amssymb]{revtex4-1}
\usepackage{times}
\usepackage{hyperref}
\usepackage{soul,xspace}
\usepackage{graphicx}
\usepackage[figure,table]{hypcap}

\hypersetup{colorlinks=true,
linkcolor=blue,
filecolor=blue,
citecolor=blue,
urlcolor=blue,
pdfauthor={YongYong Cai, Yongjun Yuan, Matthias Rosenkranz, Han Pu, Weizhu Bao},
pdftitle={Vortex patterns and the critical rotational frequency in rotating dipolar Bose-Einstein condensates},
pdfkeywords={dipolar BEC}}

\mathchardef\ordinarycolon\mathcode`\:
\mathcode`\:=\string"8000
\begingroup \catcode`\:=\active
  \gdef:{\mathrel{\mathop\ordinarycolon}}
\endgroup

\newcommand{\BEC}{BEC\xspace}
\newcommand{\GPE}{GPE\xspace}

\newcommand{\im}{\ensuremath{i}}
\newcommand{\eu}{\ensuremath{e}}
\newcommand{\bvec}[1]{\ensuremath{\boldsymbol{#1}}}
\newcommand{\bv}[1]{\ensuremath{\boldsymbol{#1}}}
\DeclareMathOperator{\erfc}{erfc}
\DeclareMathOperator{\erf}{erf}

\begin{document}
\title{Vortex patterns and the critical rotational frequency in rotating dipolar Bose-Einstein condensates}
\author{Yongyong Cai}
\affiliation{Beijing Computational Science Research Center, Haidian District, Beijing 100193, People's Republic of China}
\author{Yongjun Yuan}
\email{Corresponding Author}
 \affiliation{Key Laboratory of High Performance Computing and Stochastic Information Processing
(Ministry of Education of China), College of Mathematics and Computer Science, Hunan Normal
University, Changsha, Hunan 410081, People's Republic of China}
\author{Matthias Rosenkranz}
\affiliation{Department of Mathematics, National University of
  Singapore, 119076, Singapore}
  \author{Han Pu}
  \affiliation{Department of Physics and Astronomy, and Rice Quantum Institute, Rice University, Houston, TX 77251, USA}
\author{Weizhu Bao}
\affiliation{Department of Mathematics, National University of
  Singapore, 119076, Singapore}


\date{\today}

\begin{abstract}
Based on the two-dimensional mean-field equations for pancake-shaped dipolar Bose-Einstein condensates in a rotating frame with both attractive and repulsive dipole-dipole interaction (DDI) as well as arbitrary polarization angle, we study the profiles of the single vortex state and show how the critical rotational frequency change with the s-wave contact interaction strengths, DDI strengths and the  polarization angles. In addition, we find numerically that at the `magic angle'  $\vartheta=\arccos(\sqrt{3}/3)$, the critical rotational frequency is almost independent of the DDI strength. By numerically solving the dipolar GPE at high rotational speed, we identify different patterns of vortex lattices  which strongly depend on the polarization direction. As a result,  we undergo a study of vortex lattice structures  for the whole regime of  polarization direction and find evidence that the vortex lattice orientation tends to be aligned with the  direction of the dipoles.
\end{abstract}
\maketitle

\section{Introduction}
One of the striking features of rotating atomic Bose-Einstein condensates (BECs) is the formation of
vortices above a critical angular velocity~\cite{Leggett01,Fetter09,Saarikoski10}.  In a symmetric BEC,
multiple vortices arrange in a characteristic triangular pattern
~\cite{Fetter09}.  This triangular vortex lattice minimizes the
free energy of the BEC.

While the initial experiments considered atoms with local interactions, more recently, dipolar BECs with electric or magnetic dipole moment have received much attention from both theoretical and experimental studies (for recent reviews, see
Refs.~\cite{LahMenSan09,Bar08}). Their dipole-dipole interaction (DDI) crucially affects the
ground-state properties~\cite{GorRzaPfa00,YiYou00},
stability~\cite{SanShlZol00,SanShlLew03,Fis06,ODell2007}, and dynamics of the
gas~\cite{YiYou01}.  Furthermore, they offer a route for studying
 many-body quantum effects, such as a superfluid-to-crystal
quantum phase transition~\cite{BueDemLuk07},
supersolids~\cite{GorSanLew02} or even topological
order~\cite{MicBreZol06}.  Recent advances in experimental techniques
have paved the way for a Bose-Einstein condensate (BEC) of
\textsuperscript{52}Cr with a magnetic dipole moment $6\mu_B$ (Bohr
magneton $\mu_B$), much larger than conventional alkali
BECs~\cite{GriWerHen05,StuGriKoc05,KocLahMet08}.  Promising candidates
for  dipolar BEC experiments are Er and Dy with even larger
magnetic moments of $7\mu_B$ and $10\mu_B$,
respectively, which have been reported in experiments ~\cite{Aikawa2012,LuLev2011}.  Furthermore, DDI-induced
decoherence and spin textures have been observed in alkali-metal
condensates~\cite{FatRoaDei08,VenLesGuz08}.  Dipolar effects also play
a crucial role in experiments with Rydberg atoms~\cite{VogVitZha06}
and heteronuclear molecules~\cite{NiOspWan10,DeMChoNey11}.  Bosonic
heteronuclear molecules may provide a basis for future experiments on
BECs with dipole moments much larger than  in atomic
BECs~\cite{VoiTagCos09}.

The anisotropy of the dipole-dipole interactions crucially affects
stationary states of the rotating dipolar BEC.  In this
article we focus on a system of dipolar BEC confined in a quasi-two-dimensional pancake shaped trapping potential with the atomic magnetic dipoles polarized by an external magnetic
field. We define the polarization angle $\vartheta$ to be the angle between the dipoles and the direction normal to the plane of the condensate. Hence, if the dipoles lie in the plane of the condensate, we have $\vartheta =\pi/2$; whereas if the dipoles are perpendicular to the plane, we have $\vartheta =0$. By adjusting the external magnetic field, $\vartheta$ can be varied smoothly between 0 and $\pi/2$. Most previous studies of
rotating dipolar BECs focused only on the two limiting cases with $\vartheta =0$ or $\pi/2$~\cite{Cooper05,Cooper06,ZhangZai05,HanPu06,Zhaoan10}.
Recently,  \citet{ZhaoGu16} and \citet{MalKriRei11} studied the angular momentum and
critical rotational frequency of a 2D dipolar BEC with positive DDI strength in this intermediate
regime.
Their results show that the critical rotational frequency increases with the polarization angle $\vartheta$, while the relation between the critical rotational velocity and the DDI strength is dependent of the polarization angles.  \citet{Martin17} analytically studied the vortex lattice for the case where the dipoles are not perpendicular to the plane of rotation, and suggested there is a phase transition in the vortex geometry from triangular to square which can be measured as a function of the DDI strength,  whereas the vortex orientation is independent of the polarization angle $\vartheta$. This vortex structure transition was observed in the numerical results of \citet{ZhaoGu16} for a rotating quasi-2D dipolar BEC with positive DDI strength, however, to our knowledge, there have not,  to date, been numerical results concerning  the change of vortex lattice orientation with the polarization angle $\vartheta$. In this paper, we further study the effect of the s-wave contact interaction strength and the polarization angle on the critical rotational frequency with both attractive and repulsive DDI, and focus on fast rotation with many
vortices.  We observe
different patterns of vortex lattices which strongly depend on the
polarization direction and illustrate the vortex orientation by virtue of the static structure
factor~\cite{ReiOlsSca01,PuBakYi05}. We also take into account attractive DDI
which can be achieved with a rotating magnetic
field~\cite{GioGoePfa02}.  Simulating high vortex numbers requires
reliable numerical methods. Spectral methods are very accurate for such kinds
of problems~\cite{BaoCaibook,BaoCaiWan10,BaoTang16,Huang10,TangZhang17}, with  less grid points needed than those of
traditional finite difference methods.

This article is organized as follows.  In Sec.~\ref{sec:model} we
present a 2D model for a dipolar BEC in the rotating frame. We also
explain our approach for solving this model numerically.  In Sec.~\ref{sec:frequency}, we show how the s-wave contact interaction strength and the polarization angle affect the critical rotational frequency with both attractive and repulsive DDI strengths. In
Sec.~\ref{sec:vortices} we present simulation results of
stationary states at high rotation frequency for different
polarization angles and DDI strengths.  Focusing on the regime with many
vortices allows us to discern characteristic vortex patterns that
occur as the polarization changes from predominantly perpendicular to
parallel.  We conclude in Sec.~\ref{sec:conclusion}.

%\begin{itemize}
%\item In contrast to Malet et al. we focus on the patterns in fast
%  rotating BECs in a symmetric trap.  Simulating high vortex numbers
%  requires reliable numerics etc.
%\item Show phase of vortices
%\item We observe a striking change in vortex pattern as the
%  polarization axis changes from perpendicular to parallel.
%\item We also study attractive dipole-dipole interaction
%\end{itemize}

\section{Model}\label{sec:model}
We consider a polarized dipolar \BEC trapped in a harmonic potential
$V(\bv r) = \tfrac{1}{2} m[\omega_r^2(x^2+y^2) + \omega_z^2 z^2]$ with
$m$ the atomic mass and $\omega_r$, $\omega_z$ the transverse and
axial trap frequencies, respectively.  We assume that the magnetic dipoles are polarized along an axis ${\bf n}=(\cos \varphi \sin \vartheta, \sin \varphi \sin \vartheta, \cos \vartheta)$, where $\varphi$ and $\vartheta$ are the azimuthal and polar angles, respectively. The DDI
potential between two atoms separated by the relative vector $\bvec r$
is given by
\begin{equation}
  U_\text{dd}(\bvec r) = \frac{g_d}{4\pi} \frac{1 -
    3\cos^2\theta}{|\bvec r|^3}.
\end{equation}
Here, $\theta$ is the angle between the polarization axis and $\bvec
r$. For magnetic dipoles the interaction strength $g_d$ is given by
$g_d = \mu_0\mu_d^2$, where $\mu_0$ is the magnetic vacuum
permeability and $\mu_d$ the dipole moment.  In addition, we assume
that the \BEC is rotating with frequency $\Omega$ around the $z$ axis.
In the remainder of this article we adopt length, time and energy units as $a_r = \sqrt{\hbar/m\omega_r}$,
 $1/\omega_r$, and  $\hbar\omega_r$, respectively.  At zero
temperature this system is described by the Gross-Pitaevskii equation
(\GPE) in the rotating frame~\cite{PitStr03,YiYou01}
\begin{equation}\label{eq:GPE}
  \begin{split}
    \im\partial_t \Psi(\bvec r, t) &= \biggl[-\frac{1}{2} \nabla^2 +
    V(\bvec r) - \Omega L_z + g|\Psi|^2\\
    &\quad + \int d\bvec r' U_\text{dd}(\bvec r - \bvec r')
    |\Psi(\bvec r', t)|^2 \biggr] \Psi(\bvec r, t).
  \end{split}
\end{equation}
Here, $L_z = \im(y\partial_x - x\partial_y)$ is the $z$ component of
the angular momentum operator and $g = 4\pi N a_s/a_r$ with $N$ being the
number of atoms and $a_s$ being the s-wave scattering length.  The
dimensionless DDI strength is given by $g_d = Nm \mu_0
\mu_d^2/3\hbar^2 a_r$.

We only consider the case where $\omega_z \gg \omega_r$ and the
contact and dipole-dipole interaction energies are smaller than
$\hbar\omega_z$ such that the \BEC remains in the ground state of the axial
harmonic trap.  This is the limit of a quasi-2D
\BEC~\cite{PetHolShl00}.  The wave function $\Psi(\bvec r, t)$ can be
separated into a radial and longitudinal part, that is, $\Psi(\bvec r,
t) = \psi(\bvec\rho, t) w(z) \exp(-\im \gamma t/2)$ with $\bvec\rho =
(x, y)$, $|\bvec\rho| =\sqrt{x^2+y^2}$, $w(z) = (\gamma/\pi)^{1/4} \exp(-z^2/2\gamma)$, and $\gamma =
\omega_z/\omega_r$.  Inserting this expansion of the wave function
into Eq.~\eqref{eq:GPE} and integrating out the $z$ axis reduces
Eq.~\eqref{eq:GPE} to~\cite{CaiRosLei10,YiPu06}
\begin{equation}\label{eq:psi_2d}
  \begin{split}
    \im \partial_t \psi(\bvec\rho, t) &= \biggl[ -\frac{1}{2}
    \nabla_r^2 + \frac{|\bvec\rho|^2}{2} - \Omega L_z + \bar g
    |\psi(\bvec\rho, t)|^2 \\
    &\quad + \int d\bvec\rho' U_\text{dd}^\text{2D}(\bvec\rho -
    \bvec\rho') |\psi(\bvec\rho', t)|^2 \biggr] \psi(\bvec\rho, t).
  \end{split}
\end{equation}
Here, $\nabla_r^2 = \partial_x^2 + \partial_y^2$ and $\bar g =
\sqrt{\tfrac{\gamma}{2\pi}} \left[g - g_d \left(1 - 3 \cos^2\vartheta
  \right) \right]$ is the effective 2D contact interaction that now
depends on the DDI strength and polarization direction.  The effective
kernel for the 2D DDI is given by
\begin{widetext}
\begin{equation}\label{eq:U_2d}
  \begin{split}
    U_\text{dd}^\text{2D}(\bvec\rho) &=
    \frac{g_d\gamma^{3/2}}{8\sqrt{2\pi^3}}
    \eu^{\gamma |\bvec\rho|^2/4} \Bigl[ \left(1 - 3\cos^2\vartheta +
      \gamma[(x \cos\varphi + y\sin\varphi)^2\sin^2\vartheta -
      |\bvec\rho|^2\cos^2\vartheta]\right)
    K_0(\gamma |\bvec\rho|^2/4)\\
    &\quad -\left(1 - \cos^2\vartheta + \gamma [(x\cos\varphi +
      y\sin\varphi)^2 [1 - 2/\gamma |\bvec\rho|^2] \sin^2\vartheta -
     | \bvec\rho|^2 \cos^2\vartheta] \right) K_1(\gamma| \bvec\rho|^2/4)
    \Bigr],
  \end{split}
\end{equation}
\end{widetext}
where $K_\nu$ are modified Bessel functions of the second kind.  In
Fourier space the DDI [potential in second line in
Eq.~\eqref{eq:psi_2d}] becomes $\hat V_\text{2D}(\bv k) = \hat
U_\text{2D}(\bv k) {|\hat{\psi}(\bv k)|^2}$ with $\hat{\psi}(\bv k)$ being the condensate wave function in momentum space and $\hat
U_\text{2D}(\bv k) = \tfrac{3g_d}{2} [(\hat k_x \cos\varphi + \hat k_y
\sin\varphi)^2\sin^2\vartheta - \cos^2\vartheta] k \eu^{k^2/2\gamma}
\erfc(k/\sqrt{2\gamma})$, where $k = |\bv k|$, $\hat k_{x,y} =
k_{x,y}/k$ are normalized components of the momentum, and $\erfc(x) =
1 - \erf(x)$ is the complementary error function.
%We
%note that in the laboratory frame, the polarization axis co-rotates
%with frequency $\Omega$ around the $z$ axis.

The effective nonlocal interaction of a quasi-2D dipolar \BEC,
Eq.~\eqref{eq:U_2d}, is attractive along the projection of the
polarization axis $(\cos\varphi, \sin\varphi)$ and repulsive
perpendicular to the polarization axis. For axial polarization $\vartheta=0$, the
nonlocal interaction is isotropic and repulsive. In our work, without loss of generality, we assume that the dipoles are polarized in the $xz$-plane, such that we can fix $\varphi=0$. The effective
interaction diverges less strongly in the limit $|\bvec\rho|
\rightarrow 0$ than the full 3D dipole-dipole potential $U_\text{dd}$.
Furthermore, it has a well-behaved Fourier transform, which is
advantageous for numerical computations~\cite{CaiRosLei10}. To find the ground states, we
use the imaginary time method, with backward Euler discretization in time and Fourier spectral discretization in space. In our calculation, the effective 2D nonlocal term is evaluated by Fast Fourier Transform. When the total energy attains its global minimum by the imaginary time prorogation~\cite{BaoCaibook,BaoCaiWan10,BaoTang16}, the ground state is obtained.

%with the corresponding energy given by $\frac{1}{2}\int d\bv k  \ \hat
%U_\text{2D}(\bv k) \widehat{|\psi|^2}(\bv k)$.

\section{Critical rotational frequency}\label{sec:frequency}

In this section, we show the impact of varying s-wave contact interaction strength $g$, DDI strength $g_d$ and  polarization angle $\vartheta$ to the critical rotational frequency, respectively. Actually, \citet{MalKriRei11} have studied the angular momentum and critical rotational frequency of a dipolar BEC in the intermediate
regime with positive DDI strength, here we further study it for rotating dipolar BECs with both positive and negative DDI. We are also interested in the change of single vortex shape with different polarization angle $\vartheta$.

\begin{figure}[ht!]
  \includegraphics[width=\linewidth]{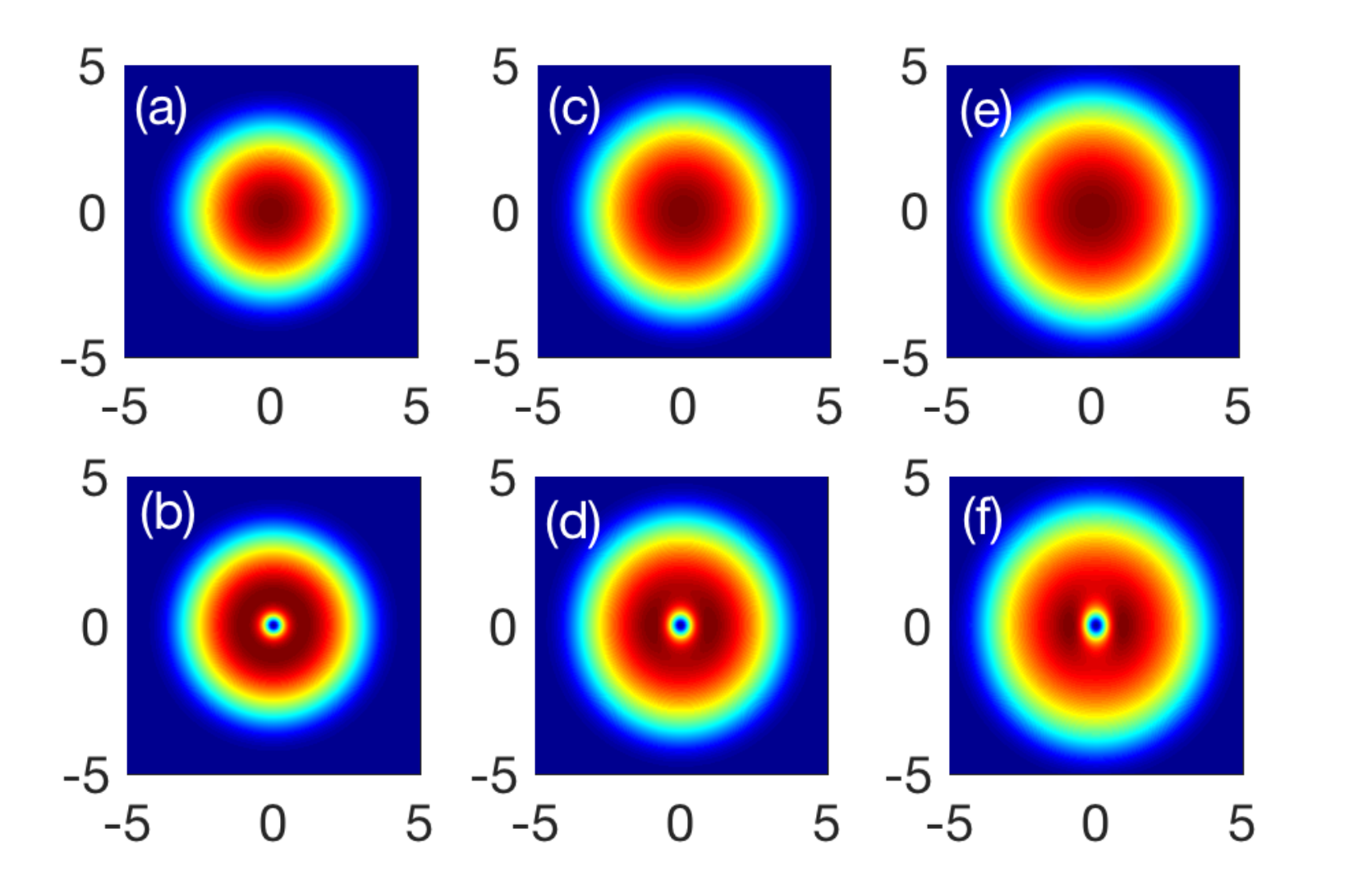}%
  \caption{(Color online) Density  of a rotating dipolar \BEC for fixed $\gamma = 10$, $g = 250$, $g_d = -100$, different polarization angles and rotation frequency.  (a) $\varphi=\vartheta= 0$ and $\Omega=0.356$. (b) $\varphi=\vartheta = 0$ and $\Omega=0.357$. (c) $\varphi=0$, $\vartheta =\pi/4$ and $\Omega=0.275$. (d) $\varphi=0$, $\vartheta =\pi/4$ and $\Omega=0.276$.  (e) $\varphi=0$, $\vartheta = \pi/2$ and $\Omega=0.236$. (f) $\varphi=0$, $\vartheta =\pi/2$ and $\Omega=0.237$.   }\label{fig:critical_omg1}
\end{figure}

\begin{figure}[ht!]
  \includegraphics[width=\linewidth]{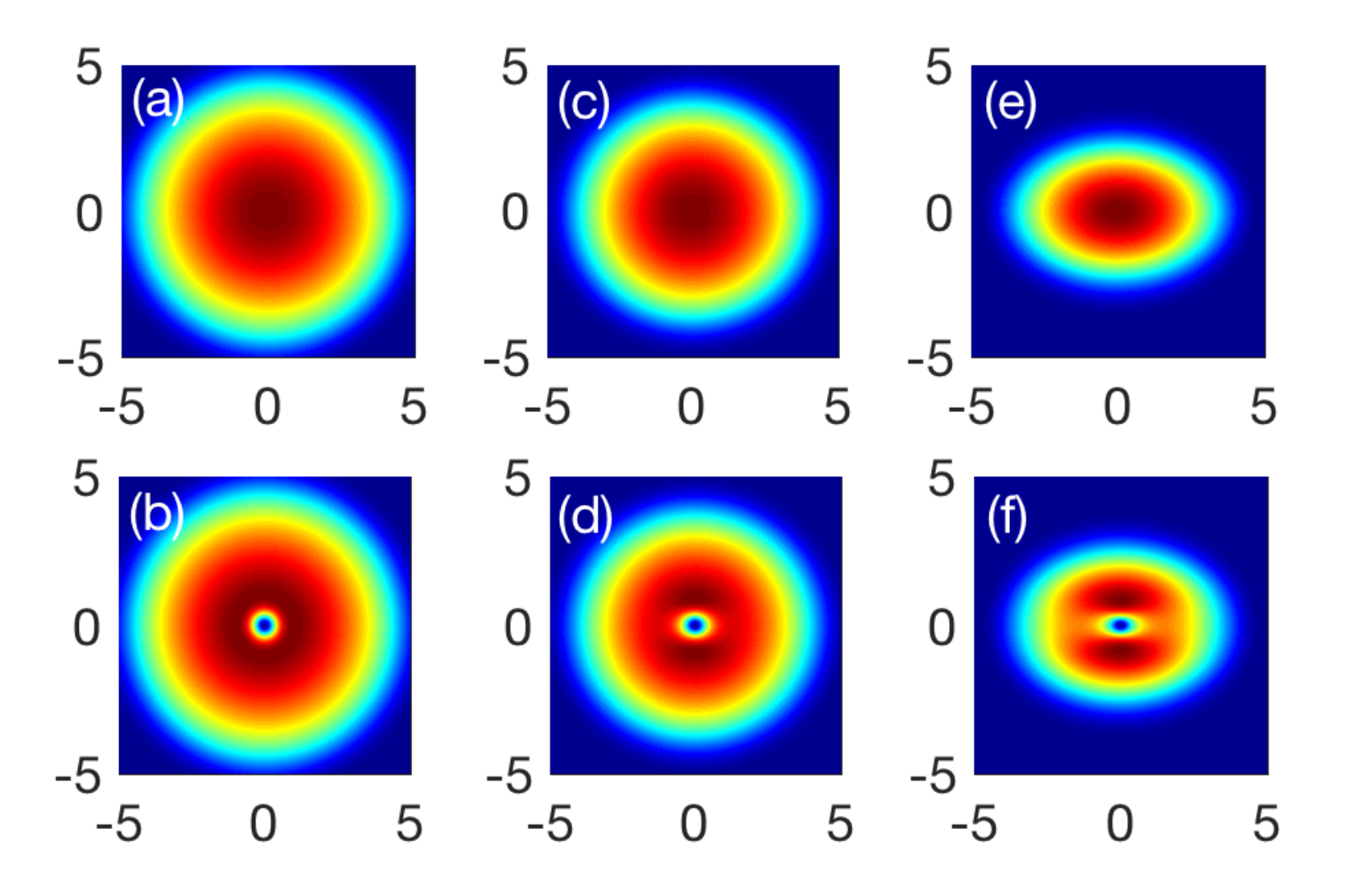}%
\caption{(Color online) Density  of a rotating dipolar \BEC for fixed $\gamma = 10$, $g = 250$, $g_d = 200$, different polarization angles and rotation frequency.  (a) $\varphi=\vartheta= 0$ and $\Omega=0.195$. (b) $\varphi=\vartheta = 0$ and $\Omega=0.196$. (c) $\varphi=0$, $\vartheta =\pi/4$ and $\Omega=0.232$. (d) $\varphi=0$, $\vartheta =\pi/4$ and $\Omega=0.233$.  (e) $\varphi=0$, $\vartheta = \pi/2$ and $\Omega=0.357$. (f) $\varphi=0$, $\vartheta =\pi/2$ and $\Omega=0.358$.   }\label{fig:critical_omg2}
\end{figure}

Figures~\ref{fig:critical_omg1}-\ref{fig:critical_omg2} show density contour plots of the rotating
dipolar \BEC for different polarization angles $\vartheta$ and rotational frequency $\Omega$ with negative and positive DDI strength $g_d$, respectively. It is observed that for the fixed effective 2D contact interaction strength $\bar g$ and the kernel $U_\text{dd}^\text{2D}(\bvec\rho)$ for the DDI in Eq.~\eqref{eq:psi_2d}, there exists a critical rotational frequency $\Omega_c$ such that there is no vortex if $\Omega <\Omega_c$ and at least one vortex if $\Omega\geq \Omega_c$ [cf.   e.g., Figs.~\ref{fig:critical_omg1}(a)--(b) and \ref{fig:critical_omg2}(e)--(f)].  Thus vortex patterns are formed after the rotational frequency exceeding its critical value.  Furthermore, the shape of the vortices change with the polarization angle $\vartheta$. It is found that the vortex core is no longer radially symmetric and turns more and more flat along the $y$-axis (the $x$-axis) for negative (positive) DDI strength $g_d$.
From the Fourier transform of $U_{\text{dd}}^{\text{2D}}$ \eqref{eq:U_2d}, we can show that the DDI interaction potential becomes more and more repulsive (attractive) in the $x$ direction when $\vartheta$ increases from $0$
to $\frac{\pi}{2}$ for positive (negative) $g_d$. As a consequence of such anisotropic property of DDI in 2D, the density of the ground state are more concentrated in the $y$ direction for positive $g_d$ and in the $x$ direction for negative $g_d$.
In contrast, when $g_d=0$ or $\vartheta=0$, the DDI in the 2D rotating dipolar \BEC Eq.~\eqref{eq:psi_2d} is isotropic in the $xy$-plane,  and the vortices in both cases possess radial symmetry in shape.

\begin{figure}[ht!]
  \includegraphics[width=\linewidth]{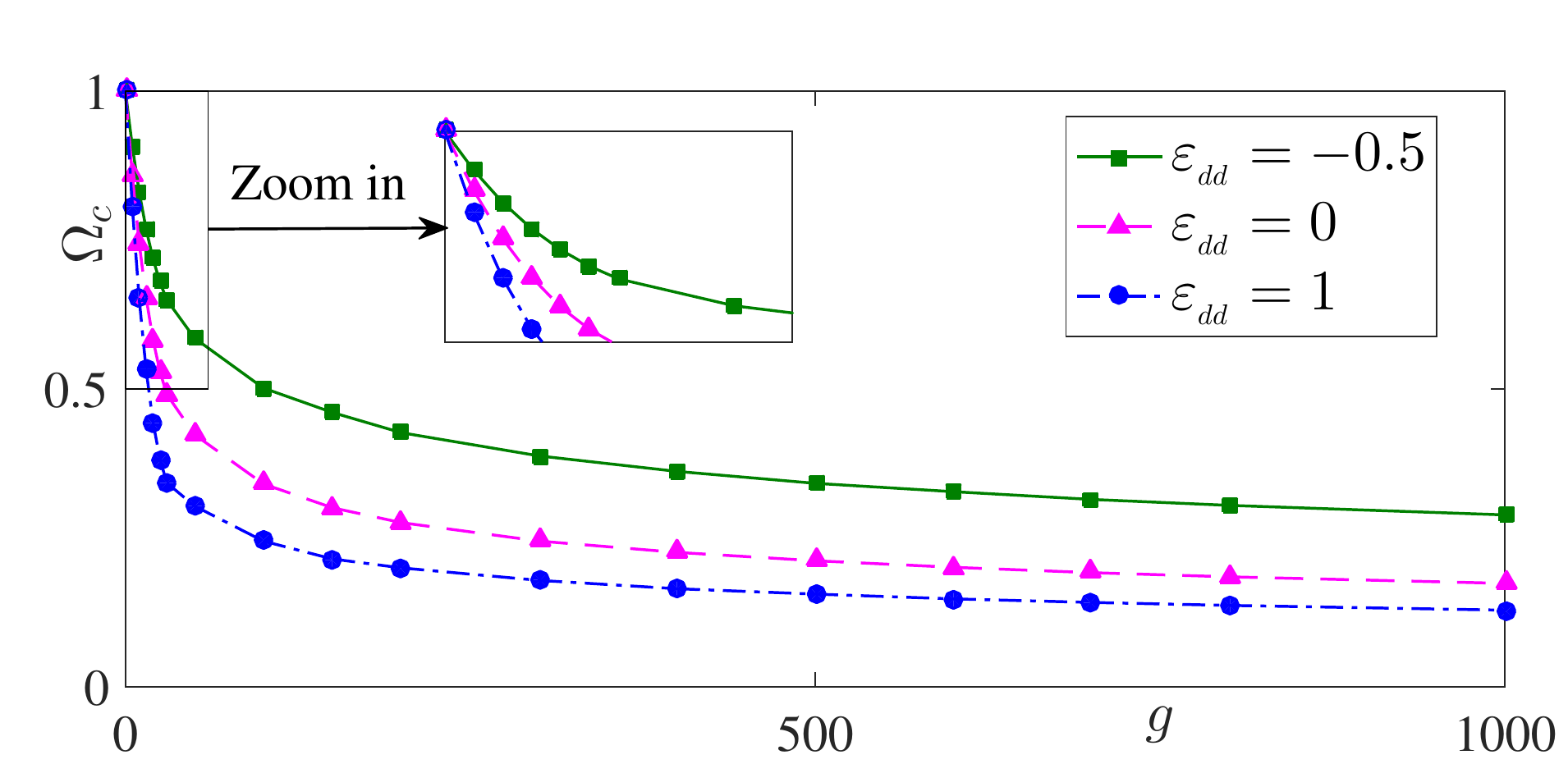}%
\caption{The change of critical rotation frequency $\Omega_c$ of a rotating dipolar \BEC with the s-wave contact interaction strength $g$ for fixed $\gamma = 10$, ${\bf n=(0,0,1)}$ and a natural dimensionless parameter $\varepsilon_{_{dd}}:= g_d/g=$ - 0.5,  0 and 1, respectively. }\label{fig:critical_omg3}
\end{figure}

Figure~\ref{fig:critical_omg3} illustrates the change of critical rotational frequency $\Omega_c$ with varying  s-wave contact interaction strength $g$, when the polarization direction is perpendicular to the rotating plane. The numerical results show that $\Omega_c$ decreases when $g$ increases for any fixed positive and negative DDI strength $g_d$. Further, $\Omega_c\to1$ as $g\to0$ and $\Omega_c$ drops dramatically when $g$ increases near the boundary  $g\approx 0$. This is in accordance with the conventional rotating condensates without DDI \cite{Leggett01,Fetter09,Saarikoski10}. It is also clear that when the DDI strength $g_d\nearrow$, $\Omega_c\searrow$  for any fixed s-wave contact interaction strength $g$ and other parameters.

\begin{figure}[ht!]
  \includegraphics[width=0.98\linewidth]{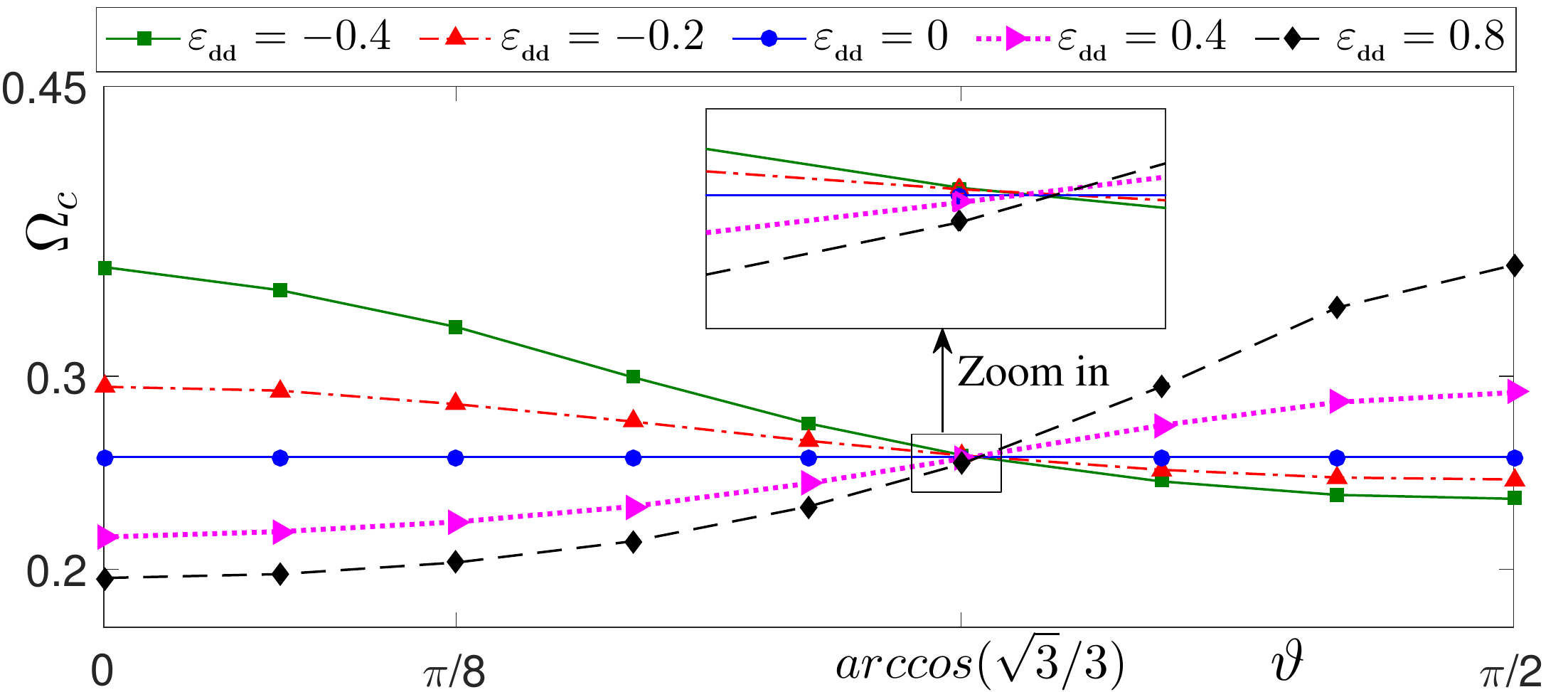}%
\caption{The change of critical rotation frequency $\Omega_c$ of a rotating dipolar \BEC with the polarization angle $\vartheta$ for fixed $\gamma = 10$, $g=250$ and $\varepsilon_{_{dd}}$= - 0.4,   - 0.2,  0,  0.4 and 0.8, respectively. }\label{fig:critical_omg4}
\end{figure}

Figure~\ref{fig:critical_omg4} shows the critical rotational frequency $\Omega_c$ versus the polarization angle $\vartheta$. The numerical results show that $\Omega_c$ decreases (increases) when $\vartheta$ varies from $0$ to $\pi/2$ for any fixed negative (positive) DDI strength $g_d$. Moreover, the curves of $\Omega_c$ as functions of $\vartheta$ with both negative and positive DDI strength $g_d$ almost intersect with each other at the `magic angle'  $\vartheta=\arccos(\sqrt{3}/3)$, since at this angle the effective 2D contact interaction in Eq.~\eqref{eq:psi_2d} is independent of the DDI strength $g_d$ and the DDI is much smaller compared to the effective 2D contact interaction thus has very little effect on the critical rotational frequency.

\section{Vortex patterns}\label{sec:vortices}
\begin{figure}[ht!]
  \includegraphics[width=\linewidth]{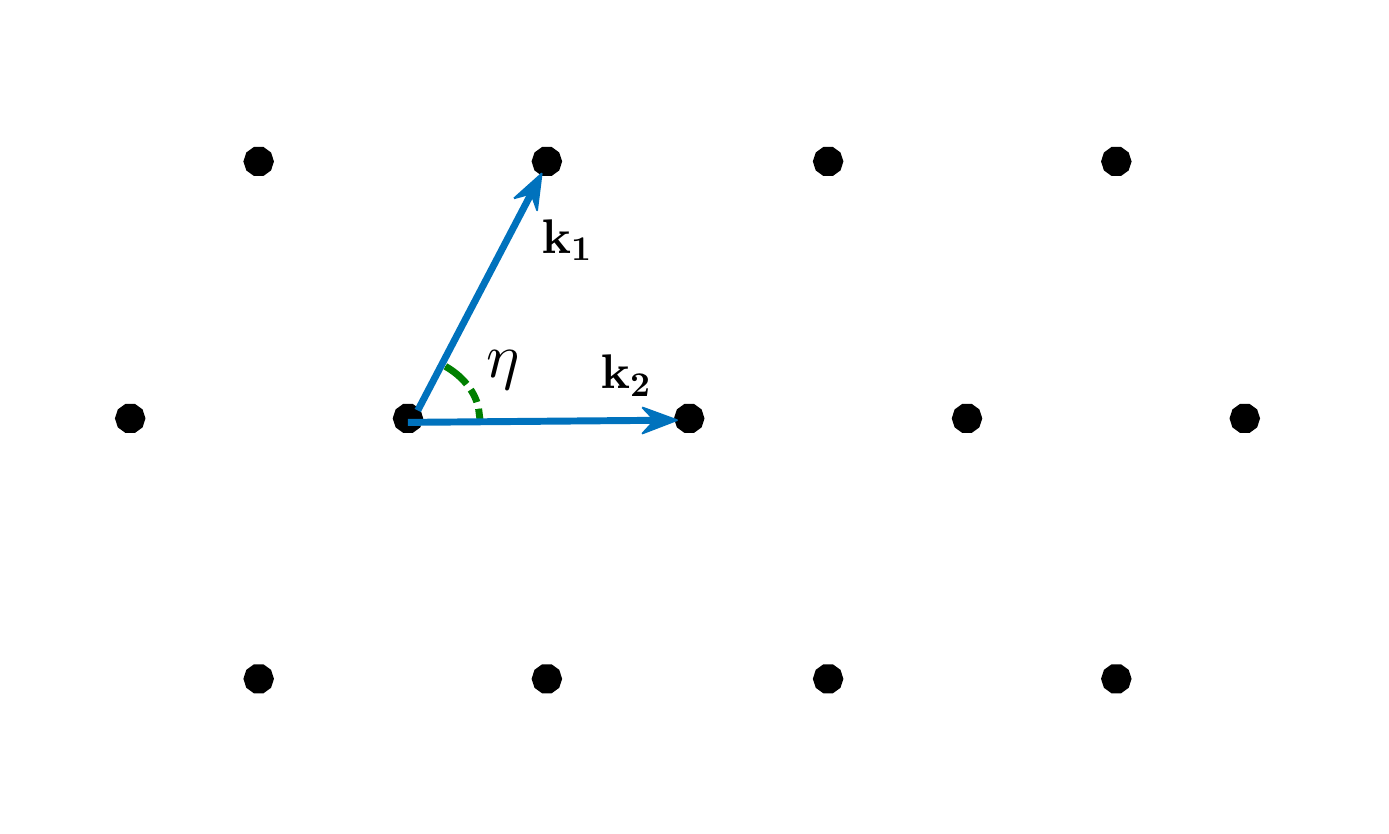}
  \caption{Illustration of the Bravais lattice basis vectors and the
lattice parameters.}\label{fig:oriention}
\end{figure}

In this section, we show different vortex lattices that
emerge as stationary states for varying polarization angles under fast rotation. To characterize
the structure of the vortex lattice, we define the static structure
factor~\cite{ReiOlsSca01,PuBakYi05}
\begin{equation}
  S(\bv k) = \frac{1}{N_v^2} \biggl|\sum_j \eu^{\im\bv k\cdot\bv\rho_j} \biggr|^2,
\end{equation}
where $N_v$ is the number of vortices and $\bv\rho_j$ are their positions.
The structure factor exhibits peaks at the reciprocal lattice sites,
which reveal the frequencies and orientation of the vortex lattice.
The reciprocal lattice is defined by two basis vectors $\bv k_1$ and
$\bv k_2$. Here  we choose $\bv k_1$ as the one closest to the $y$-axis
 and use the parameter $\eta=
\angle(\bv k_1,\bv k_2)$ to characterize the
orientation of the vortex lattice [c.f. Fig.~\ref{fig:oriention}]. $\eta = \pi/2$ for a rectangular vortex lattice, and $\pi/3$ for a trianglar lattice.

\begin{figure}[ht!]
  \includegraphics[width=\linewidth]{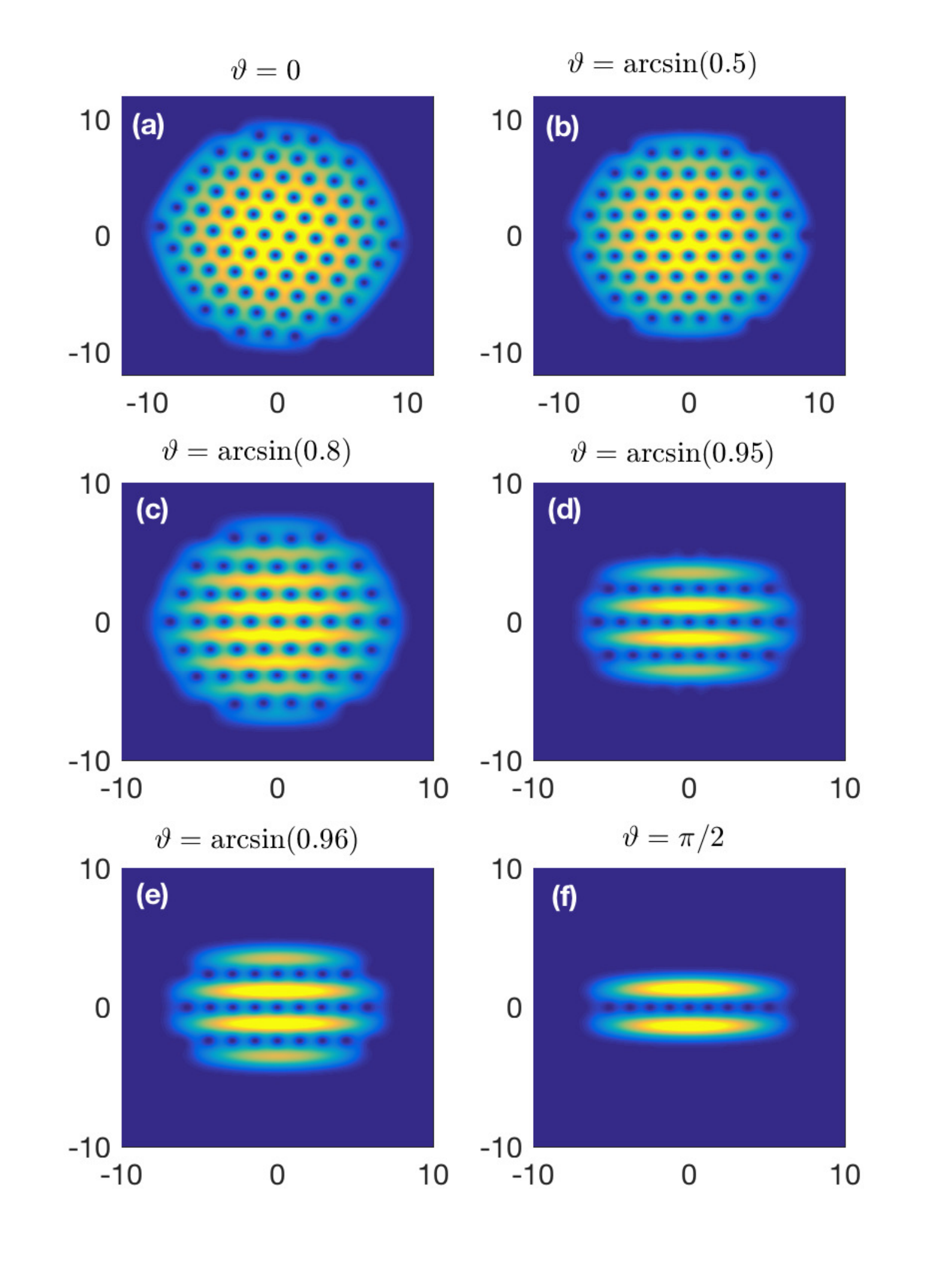}
  \caption{(Color online) Density  of a rotating dipolar \BEC for different
    polarization angles. $\varphi=0$  and $\vartheta = 0,\arcsin(0.5),\arcsin(0.8),\arcsin(0.95),\arcsin(0.96),\pi/2$ (from left to right, top to bottom).  The rotation frequency is $\Omega/\omega_r = 0.95$,
    $\gamma = 10$, $g = 250$, and $g_d = 250$.}\label{fig:patterns}
\end{figure}

\begin{figure}[ht!]
  \includegraphics[width=\linewidth]{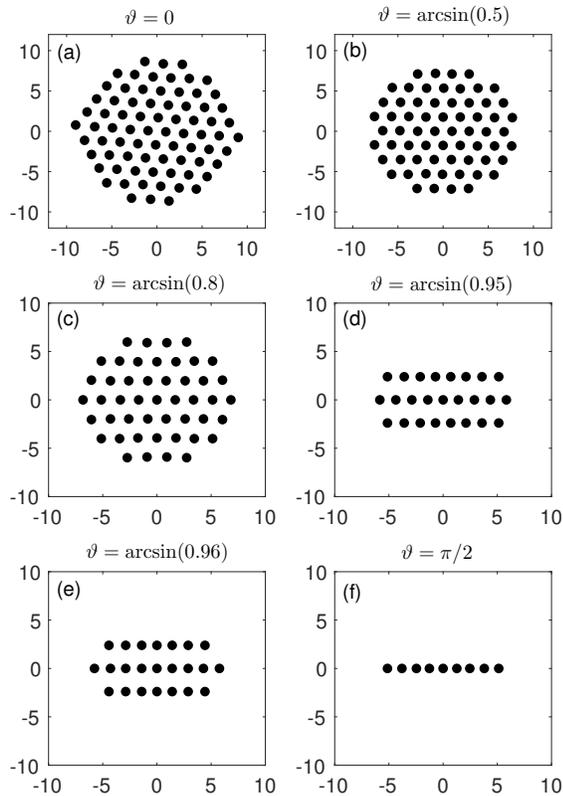}
  \caption{Vortex positions.  Same parameters as in Fig.~\ref{fig:patterns}.}\label{fig:lattice}
\end{figure}

%\subsection{Vortex lattice orientation}

\begin{figure}[ht!]
  \includegraphics[width=\linewidth]{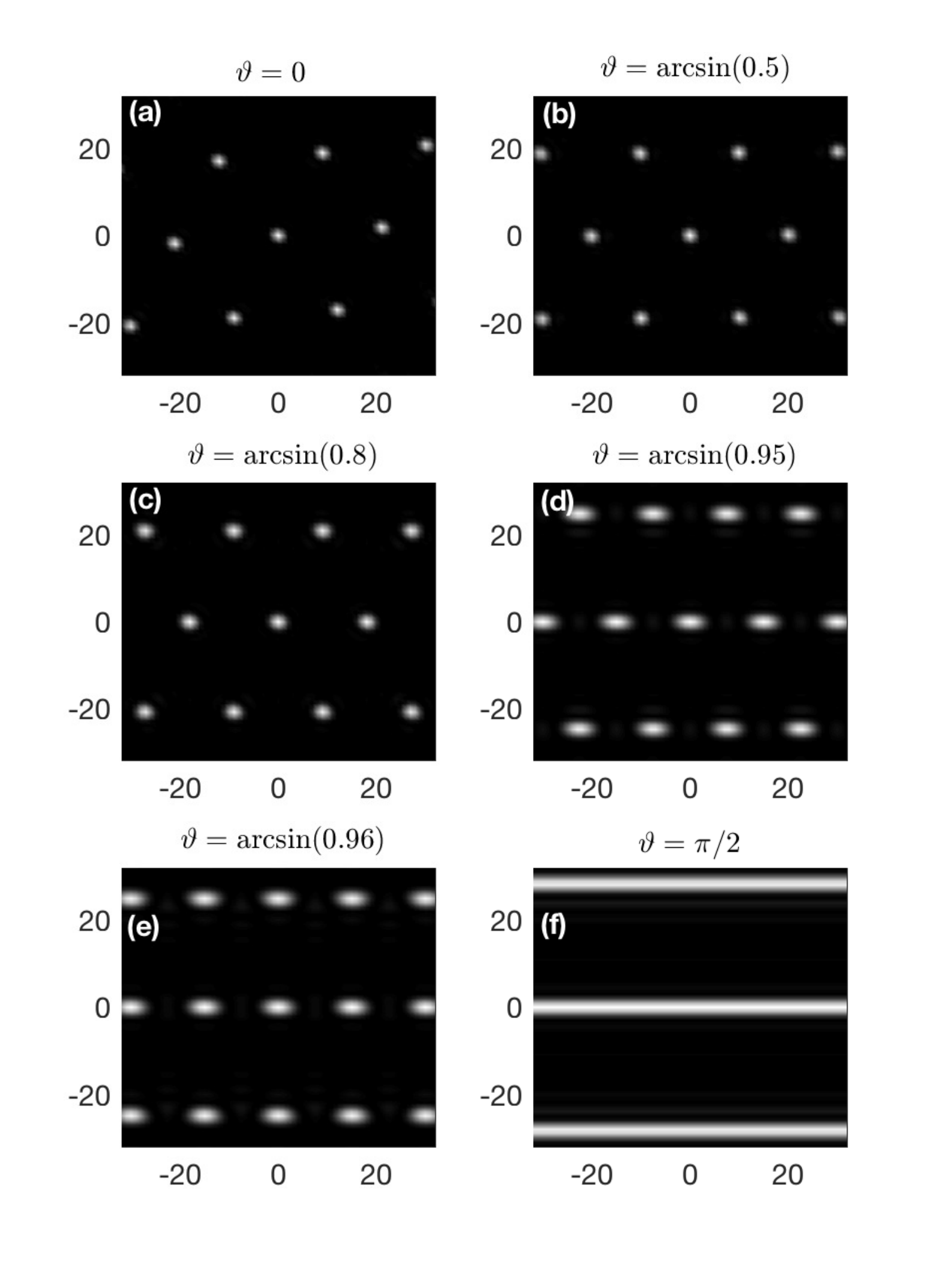}
  \caption{Static structure factor S. Same parameters as in Fig.~\ref{fig:patterns}.}\label{fig:static}
\end{figure}

We start with the impact of the polarization direction on the vortex lattice geometry under the fast rotation.
We compute the ground states of the dipolar \BEC for different polarization angles $\vartheta$ at strong
DDI $g_d = g$ and high rotation frequency $\Omega/\omega_r = 0.95$ by imaginary time propagation.
As shown in Fig.~\ref{fig:patterns}, for polarization predominantly along the symmetry axis of the BEC (i.e., $\vartheta \approx 0$), the
vortices form a regular triangular lattice
[cf. Figs.~\ref{fig:patterns}(a)--(d)].  The corresponding structure factor
in Figs.~\ref{fig:static}(a)--(d) reveals a hexagonal reciprocal primitive
cell, characteristic of the triangular lattice.  As the polarization axis
rotates into the plane of the BEC, the vortex lattice aligns with the
polarization [cf. Figs.~\ref{fig:patterns}(c)--(f)].  Parallel
polarization (i.e., $\vartheta \approx \pi/2$) is shown in Figs.~\ref{fig:patterns}(e)--(f).  In the
extreme case $\vartheta=\pi/2$, the vortices align on a central 1D lattice that splits
the BEC.  The elongation in each BEC fragment is caused by
magnetostriction, which tends to align dipoles in a head-to-tail
configuration.  The DDI between the two fragments is repulsive but
drops exponentially~\cite{BarMicRon11,RosBao11}.  The distance
between the fragments is $\simeq 2.5 a_r$, which is on the order of
$\mathrm{\mu m}$. The corresponding static structure factor in
Fig.~\ref{fig:static}(f) is nearly uniform in the perpendicular
direction and periodic along the polarization direction.  For polarization angles which are slightly less than $\pi/2$, instead of a single split we
 observe splits into several BEC fragments caused by a nearly
 rectangular vortex lattice  [cf. Figs.~\ref{fig:patterns}(e)].  In this
 configuration the BEC minimizes its energy by forming quasi-1D tubes
 between the vortices aligned along the polarization direction.  Again,
 the orientation of these tubes is dictated by magnetostriction.

Positions of vortex cores are displayed in Fig.~\ref{fig:lattice}.  If
the polarization is predominantly perpendicular to the rotating plane, the vortex lattice is
regular, as observed in the rotating BEC without DDI \cite{Leggett01,Fetter09,Saarikoski10,Du2001,baoWM2005,Zeng2009}.  For increasing polarization in the plane, it changes
orientation to align with the polarization direction.  Rotating the
polarization further into the plane, magnetostriction elongates the
BEC.  Consequently, the vortex lattice becomes more irregular as rows
of vortices perpendicular to the polarization are shaved off.

\begin{figure*}[ht!]
 \hspace{-1.0cm} \includegraphics[width=\linewidth]{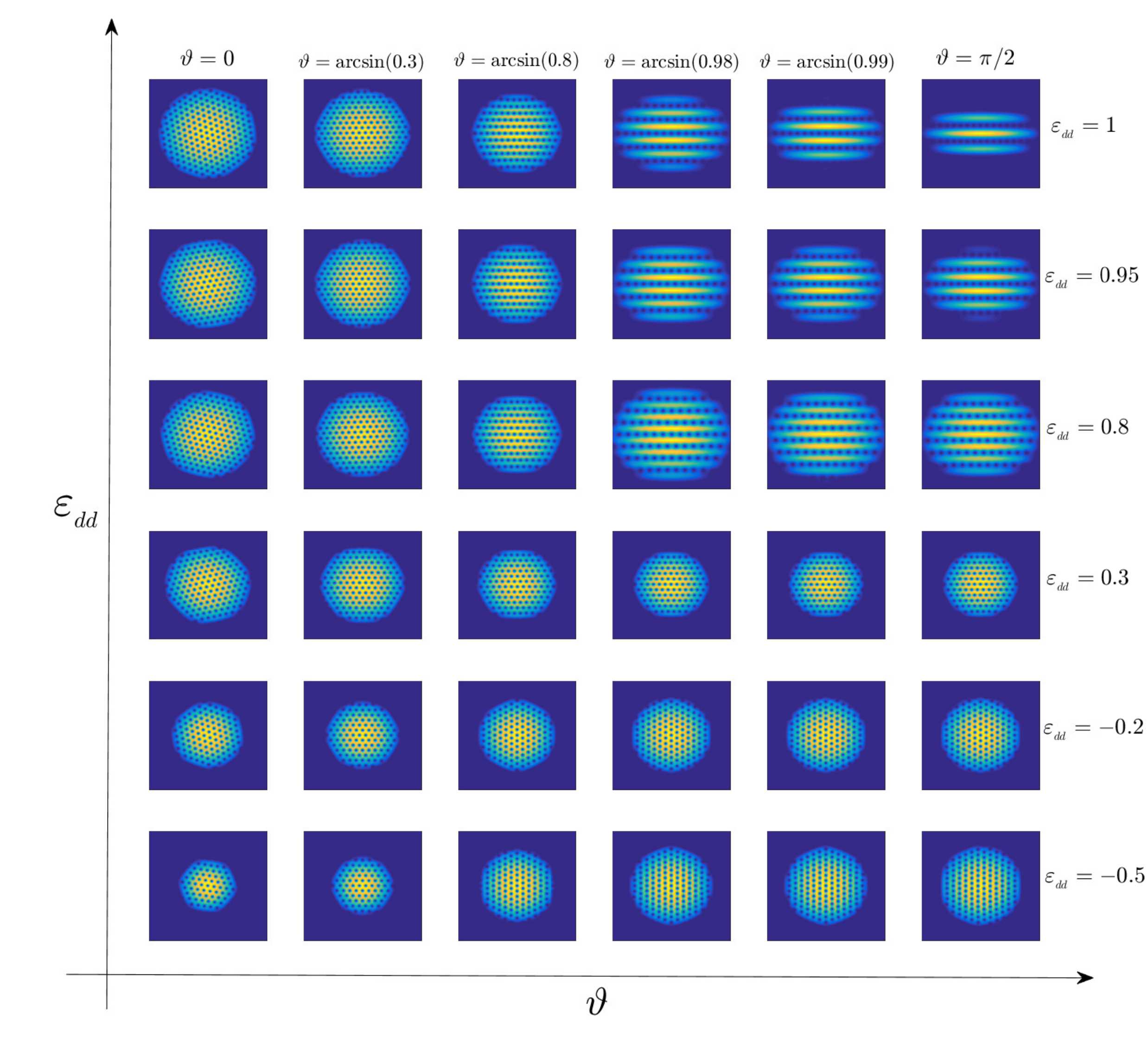}
  \caption{(Color online) Density  of a rotating dipolar \BEC for different polarization angles with $\varphi=0$ and $\vartheta = 0, \ \arcsin(0.3)$,  $\arcsin(0.8), \ \arcsin(0.98),\ \arcsin(0.99), \  \pi/2$ (from left to right) and for different DDI strengths with $\varepsilon_{_{dd}}=-0.5, -0.2, 0.3, 0.8, 0.95, 1.0$ (from bottom to top).  The rotation frequency is $\Omega/\omega_r = 0.99$,  $\gamma = 10$ and $g = 250$.}\label{fig:dd_theta_patterns}
\end{figure*}

The effective contact interaction $\bar{g}$ in 2D dipolar GPE \eqref{eq:psi_2d} changes
 while $\vartheta$ varies from $0$ to $\pi/2$. When $\varepsilon_{_\text{dd}}>0$, $\bar{g}$ is decreasing;
 when $\varepsilon_{_{\text{dd}}}<0$, $\bar{g}$ is increasing. Based on the conventional rotating BEC results,
 the number of vortices increases when $\bar{g}$ becomes larger, and Fig.~\ref{fig:patterns} suggests it is also the case for
 BEC with DDI.
 In Fig.~\ref{fig:dd_theta_patterns},  we show density contour plots of the rotating
dipolar \BEC for different polarization angles $\vartheta$ and different DDI strength with a very fast rotation frequency $\Omega/\omega_r = 0.99$, which nearly equals to its ultimate limit $\Omega/\omega_r = 1.0$. It is observed that the change of a triangular vortex lattice structure to a rectangular vortex lattice structure occurs when both the polarization angle $\vartheta$ and the natural dimensionless parameter $\varepsilon_{_{dd}}:= g_d/g$ are close to their limits $\vartheta=\pi/2$ and $\varepsilon_{_{dd}}=1.0$. We also found that for dipoles oriented along the
$z$-axis, the system has rotational symmetry. Therefore, the patterns
plotted in Fig.~\ref{fig:dd_theta_patterns} with $\vartheta=0$ can be rotated about the $z$-axis by an
arbitrary angle, while for the other dipole orientations in the $xz$-plane, the
system has a symmetry axis along the $y$-axis, thus the vortex
lattices shown in Fig.~\ref{fig:dd_theta_patterns} with $\vartheta\in (0,\pi/2]$ all obey this symmetry.  For the negative DDI strength, there are more vortices found in the condensate for in-plane polarization of the DDI rather than off-plane polarizations, which is in contrast with
the positive DDI strength case but agrees well with the behaviour of effective contact interaction $\bar{g}$. Figs.~\ref{fig:patterns} and \ref{fig:dd_theta_patterns}
imply that the number of vortices are still mainly determined by the effective contact interaction $\bar{g}$.
On the other hand, the DDI in-plane significantly affects the distribution of the vortices (cf. Figs.~\ref{fig:static} and \ref{fig:dd_theta_patterns}). When switching to a negative DDI in Fig.~\ref{fig:dd_theta_patterns}, we find that in the extreme case $\vartheta=\pi/2$, the shapes of the vortex  lattices are much different, i.e., the major axis of the lattice with negative DDI strengths is perpendicular to the polarization $x$ axis, while the major axis of the lattice with positive DDI strengths is parallel to the polarization $x$ axis.

\begin{figure}[ht!]
  \includegraphics[width=\linewidth]{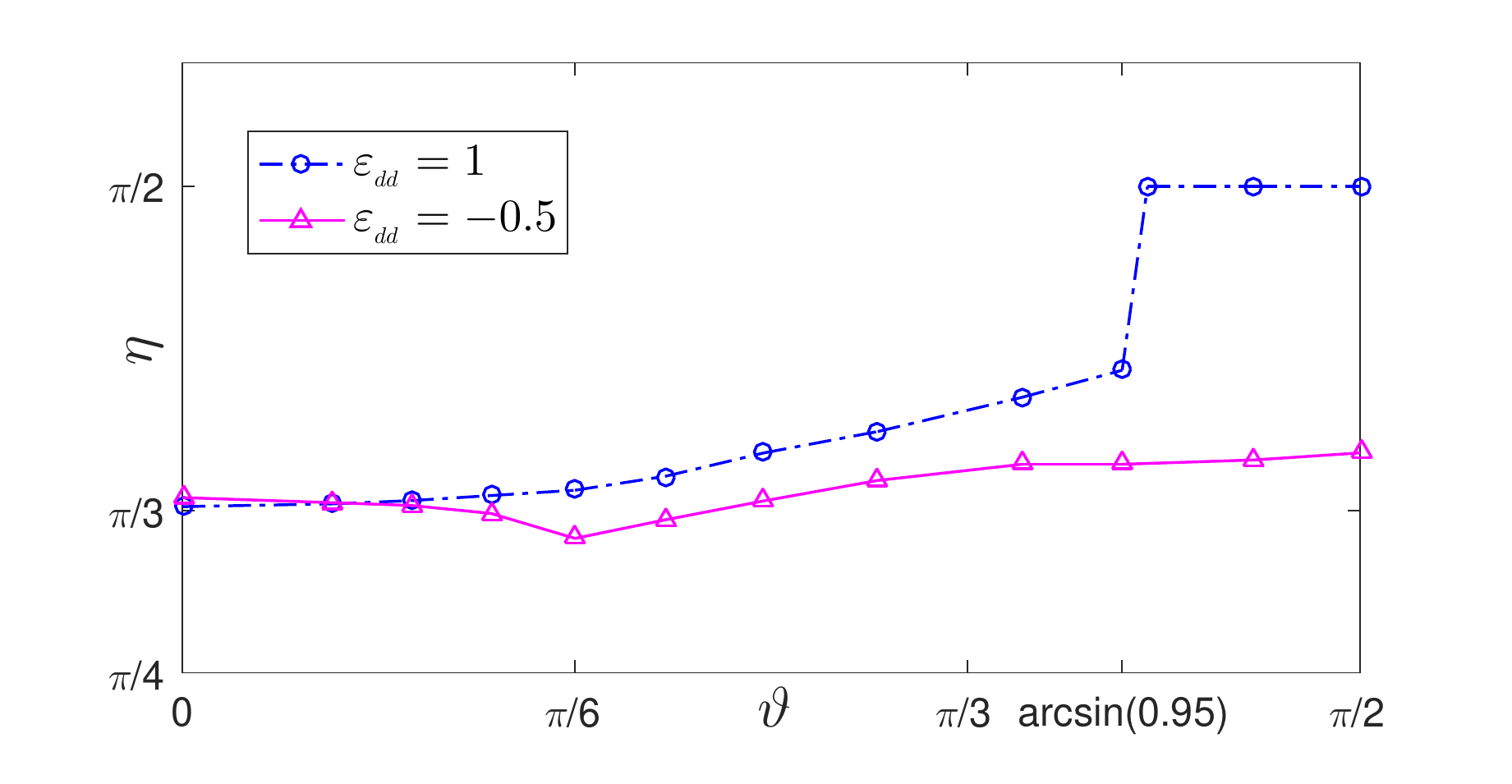}
   \caption{Lattice orientation parameter $\eta$ vs $\vartheta$. Same parameters as in Fig.~\ref{fig:patterns}.}\label{fig:eta}
\end{figure}

 We use the  parameter
$\eta$ as a function of the polarization angle $\vartheta$, as shown in Fig.~\ref{fig:eta}, to characterize the structural change of the pattern of the vortex lattice for both positive and negative DDI interaction strength. For positive DDI strength, as $\vartheta$ increases from $0$, $\eta$ starts from $\pi/3$ and varies rather slowly initially. However, at a critical angle around $\arcsin (0.95)$, $\eta$ exhibits a jump to the value of $\pi/2$, indicating a structural change to a rectangular vortex lattice. By contrast, for negative DDI strength, $\eta$ stays near $\pi/3$ as $\vartheta$ changes from 0 to $\pi/2$, hence the vortex lattice remains roughly triangular independent of the polarization angle in this case.

%\begin{figure}[ht!]
%  \includegraphics[width=0.92\linewidth]{density100}%
%  \caption{(Color online) Density  of a rotating dipolar \BEC for different
%    polarization angles. Same parameters as in Fig.~\ref{fig:patterns} except the DDI
%    parameter $g_d=100$.}\label{fig:patterns2}
%\end{figure}

%\begin{figure}[ht!]
%  \includegraphics[width=\linewidth]{density-125}%
%  \caption{(Color online) Density  of a rotating dipolar \BEC for different
%    polarization angles. Same parameters as in Fig.~\ref{fig:patterns} except the DDI
%    parameter $g_d=-125$.}\label{fig:patterns2}
%\end{figure}
%\begin{figure}[ht!]
%  \includegraphics[width=\linewidth]{S-125}
%  \caption{Static structure factor S. Same parameters as in Fig.~\ref{fig:patterns2}.}\label{fig:static2}
%\end{figure}
%\begin{figure}[ht!]
%  \includegraphics[width=\linewidth]{kappa_theta}
%  \caption{Lattice orientation parameter $\kappa$ vs $\vartheta$. Same parameters as in Fig.~\ref{fig:patterns}.}\label{fig:kappa}
%\end{figure}

%\begin{figure}[ht!]
%  \includegraphics[width=\linewidth]{phase}
%  \caption{Phases.  Same parameters as in Fig.~\ref{fig:patterns}.}\label{fig:phase}
%\end{figure}

\section{Conclusions}\label{sec:conclusion}
We have studied the change of critical rotation frequency versus the s-wave contact interaction strengths, DDI strengths and the varying polarization angles. We find that the critical rotation frequency is monotonously decreasing with growing s-wave contact interaction strength $g$, and identically approaches the confinement frequency limit for $g\approx0$ and arbitrary DDI strength $g_d$. The critical rotation frequency changes very rapidly  near $g=0$ and then decreases more and more slowly for large $g$.  In contrast to previous works,  our results cover both the case of $g_d>0$ and $g_d<0$, thus it is observed that the effect of the polarization angle $\vartheta$ to the critical rotation frequency is dependent of the sign of $g_d$, i.e., the critical rotation frequency increases (decreases) with $\vartheta$ from $0$ to $\pi/2$ for fixed positive (negative) $g_d$.  In addition, we find numerically that at the `magic angle'  $\vartheta=\arccos(\sqrt{3}/3) \approx 54.7^{\ \text{o}}$, the critical rotational frequency is almost independent of the value of DDI strength.

We have numerically simulated the dipolar GPE with fast rotation and show different patterns of vortex lattices which strongly depend on the polarization direction. When the polarization angle $\vartheta$ changes from perpendicular to parallel to the condensate plane, a structural phase transition in the vortex geometry from triangular to square is observed for positive $g_d$, but not for negative $g_d$. This result is consistence with the analytical results of \citet{Martin17}. Meanwhile, by plotting the static structure factor and the orientation parameter $\eta$ of the vortex lattice, we find evidence that the lattice orientation varies with the polarization angle $\vartheta$. Particularly when $\vartheta=\pi/2$,  a vortex lattice which is nearly uniform in the perpendicular direction and periodic along the polarization direction is obtained.

\begin{acknowledgments}
  This work was partially supported by the National Natural Science Foundation of China  Grants 11771036  and  U1530401 (Y.C.),  11601148 (Y.Y.),
  the Academic Research Fund of Ministry of Education of
  Singapore grant R-146-000-223-112 (M.R. and W.B.) and  the
  US NSF and the Welch Foundation Grant No. C-1669 (H.P.).
\end{acknowledgments}

\bibliography{diprot}

\end{document}